\documentstyle[prl,aps,multicol,epsf]{revtex}
\renewcommand{\narrowtext}{\begin{multicols}{2} \global\columnwidth20.5pc}
\renewcommand{\widetext}{\end{multicols} \global\columnwidth42.5pc} 

\begin{document}

\newcommand{\be}{\begin{equation}}
\newcommand{\ee}{\end{equation}}
\newcommand{\bea}{\begin{eqnarray}}
\newcommand{\eea}{\end{eqnarray}}
\newcommand{\nt}{\narrowtext}
\newcommand{\wt}{\widetext}

\title{Gauge (non-)invariant Green functions of Dirac fermions coupled to gauge fields}

\author{D. V. Khveshchenko}

\address{Department of Physics and Astronomy, University of North
Carolina, Chapel Hill, NC 27599}
\maketitle

\begin{abstract}
We develop a unified approach to both infrared and ultraviolet asymptotics
of the fermion Green functions 
in the condensed matter systems that allow for an effective description in the framework
of the Quantum Electrodynamics. 
By applying a path integral representation to the previously suggested 
form of the physical electron propagator we demonstrate that in the massless case
this gauge invariant function features a "stronger-than-a-pole" branch-cut singularity 
instead of the conjectured Luttinger-like behavior.
The obtained results alert one to the possibility that construction of   
physically relevant amplitudes in the effective gauge theories
might prove more complex than previously thought.
\end{abstract}

\section{Introduction}
In a generic many-body fermion system, a 
repulsive electron-electron interaction is normally expected
to result in a suppression of any amplitude which describes propagation 
of fermionic quasiparticles. For instance, in the phenomenological Fermi liquid theory,  
the residue of the electron Green function $G(\epsilon, {\vec p})=Z(\epsilon)/
(\epsilon - E({\vec p})+\mu)$
gets reduced compared to the non-interacting value ($Z(\epsilon)=1$), thus exhibiting
a partial ($0<Z(0)<1$) 
suppression of the simple pole which corresponds to the bare fermionic quasiparticles.

The question as to whether or not the repulsive fermion interactions can result in 
an even more severe, complete, 
destruction of the pole ($Z(0)=0$) remains a subject of an ongoing debate.
Such a behavior is well known to occur in the one-dimensional (1D) Luttinger 
and related models with short-ranged interactions, in which case 
the residue of the fermion Green function exhibits a characteristic 
algebraic decay $Z(p)\sim p^{\eta}$ as a function 
of the Lorentz-invariant momentum $p={\sqrt {-p^2}}=({p}^2-\omega^2)^{1/2}$
and is controlled by an anomalous dimension $\eta >0$.

In the 1D coordinate space,
this behavior corresponds to the suppression of the electron propagator
$G(t, x)\sim \sum_{\pm}\exp(\pm i{k}_F{x})/|x\pm t|^{1+\eta}$ 
which at long times and distances decays faster than the non-interacting one
($\eta=0$).
In the absence of spin, the above Green function is Lorentz invariant,
apart from the oscillating factors $\exp(\pm ik_Fx)$ that stem 
from a finite ($2k_F$) separation between the two 1D Fermi points,
in accordance with the fact that the low-energy excitations $\psi_{R,L}$ 
confined to the vicinity of the Fermi points constitute one 
Dirac fermion $\Psi=(\psi_R, \psi_L)$. 

The marked difference between this, so-called Luttinger, behavior 
and the Fermi liquid one prompts fundamentally 
important questions pertaining to the possibility of a similar behavior 
in $D>1$ and/or in the presence of long-ranged electron-electron interactions.
While in the case of the short-ranged interactions 
the possibility of the $D>1$ Luttinger-like behavior is likely to be limited 
to the infinitely strong coupling limit, the long-ranged forces  
appear to be capable of destroying the Fermi liquid even at finite couplings.
As the best studied example of this kind, the model of degenerate non-relativistic 
massive fermions 
($T\ll \mu\ll mc^2$) which are minimally coupled to an abelian gauge field  
was found to have a distinctly non-Fermi liquid behavior \cite{Reizer}, 
although the latter appears to be quite different from the Luttinger one \cite{AIM}.

More recently, there has been an upsurge of interest in the relativistic counterpart
of this model which is a zero-density ($\mu=0$) system of the $N$-flavored
relativistic Dirac fermions coupled to an  
abelian gauge field which is described by the standard action
of the Quantum Electrodynamics ($QED$)   
\be
S[\Psi,{\overline \Psi},{\bf A}]=
\int d{\bf x}[\sum_{f=1}^N{\overline \Psi}_f(i\gamma_\mu \partial_\mu+\gamma_\mu A_\mu-m)
\Psi_f+{1\over 2g^2}(\partial_\mu A_\nu - \partial_\nu A_\mu)^2]
\ee
where, for the sake of completeness, we also included a finite fermion mass $m$.
 
Among the previously discussed examples of the  
2D condensed matter systems that support the Dirac-like 
low-energy excitations and allow for such an effective description are 
the so-called flux phase in the planar quantum disordered magnets \cite{Lee,Wen}
and the layered disordered $d$-wave superconductors with strong phase fluctuations 
proposed as an explanation of 
the pseudogap \cite{Franz,Ye} and insulating (spin density wave) \cite{Herbut}   
phases of the high-$T_c$ cuprates. Also, the non-Lorentz-invariant version
of $QED_{2+1}$ was shown to provide a convenient description of the normal 
semimetalic state of highly oriented pyrolitic graphite \cite{graphite,DVK}.

The number of the fermion flavors $N$ depends on the problem in question,
although it is not necessarily equal to
the number of different conical Dirac points in the bare electron dispersion of a lattice
system. In all of the previously discussed 2D examples \cite{Lee,Wen,Franz,Ye,Herbut,graphite,DVK}, 
$N=2$ is a number of the electron spin components, while the 
number of the conical points turns out to be either two \cite{graphite,DVK}
or four \cite{Wen,Franz,Ye}) which 
merely forces one to use the four-component Dirac fermions
and the corresponding (reducible) representation of the $\gamma$-matrices
$\gamma_\mu=\sigma_\mu\otimes\sigma_3$ constructed from the triplet $\sigma_\mu$
of the Pauli matrices.

In the abovementioned 
condensed matter-related applications, the effective gauge fields serve as 
a somewhat exotic, yet often more convenient, 
representation of such bosonic collective excitations as spin 
or pairing fluctuations, while the Dirac fermions correspond to the 
auxiliary fermionic excitations such as, e.g., spinons \cite{Lee,Wen},
"topological" fermions \cite{Franz,Ye,Herbut}, and so forth. 
Generically, the quantum mechanical amplitudes describing such degrees of freedom 
turn out to be gauge-dependent, while all the physical observables which 
experimental probes can only couple to must be manifestly gauge-invariant. 

Among such gauge-invariant amplitudes, is the one containing a phase factor
(sometimes referred to as a "gauge connector" or a "parallel transporter")   
\be
G^{\Gamma}_{inv}(x, y)=<\Psi(x)e^{i\int_\Gamma A_\mu dz_{\mu}}{\overline \Psi}(y)>
\label{Ginv}
\ee
whose suggestive form makes it tempting to identify Eq.(2)
with the physical electron Green function
(in spite of its being gauge-independent, the function
$G^{\Gamma}_{inv}$ explicitly depends on the choice of the contour $\Gamma$).

To this end, it was conjectured \cite{Wen}
that by analogy with the problem of the
compressible Quantum Hall Effect described by yet another kind of 
the 2D auxiliary (this time, non-relativistic) fermionic quasiparticles, the so-called   
composite fermions, interacting with the statistical Chern-Simons field \cite{Kim}, 
the electron Green function is given by Eq.(2) with the contour $\Gamma$ 
chosen as a straight line between the ending points $\bf x$ and $\bf y$. 

Furthermore, it was argued in Ref.\cite{Wen} that in the case $m=0$
and at energies and momenta which are small as compared to the 
bandwidth and the inverse lattice spacing, respectively,
the gauge-invariant amplitude (2)
features the Luttinger-like behavior with a positive exponent $\eta$
(hereafter we use notations ${\bf q}{\bf p}=q_\mu p_\mu$ and ${\hat p}=\gamma_\mu p_\mu$)
\be
G^{|}_{inv}(p)\sim {\hat p}/p^{2-\eta}
\ee
which was also invoked in \cite{Wen,Ye}
to explain the experimental data on angular-resolved photoemission spectra 
(ARPES) in the high-$T_c$ cuprates \cite{ARPES}.

In the general case of a D-dimensional condensed matter system which possesses
a number of isolated Fermi points located at ${\vec k}_{Fi}$,
the conjectured behavior (3) 
corresponds to the algebraic suppression of the electron propagator 
at long times and distances 
\be
G^{|}_{inv}(x)\sim \sum_i e^{i{\vec k}_{Fi}{\vec x}}{{\hat x}\over x^{D+1+\eta}}
\ee
where the sum is taken over all the Fermi points. 

In the present paper, we employ a 
functional integral technique to compute the function (3) and 
discern the true nature of its singular behavior (if any). 
This approach which had been pioneered by Schwinger and later advanced by a number of 
other authors (see, e.g., \cite{Lebedev,Stefanis} and references therein) exploits a functional  
integral representation of the exact solution of the equation for  
$G^{|}_{inv}(x, y| {\bf A})$ as a functional of an arbitrary 
configuration of the gauge field ${\bf A}(z)$.
Subsequently, by averaging over the gauge field, one obtains a sum of  
all the multi-loop diagrams with no 
couplings between the fermion polarization insertions into the gauge field propagators
and the open fermion line corresponding to the fermion's propagation between the 
space-time points $\bf x$ and $\bf y$.
Likewise, in the case of a generic
multi-fermion amplitude, the allowed graphs can only contain
open fermion lines which connect the incoming and outgoing asymptotical fermionic
states, provided that the fermion polarization
has already been absorbed into the gauge field propagator.

This approach can be viewed as a systematic improvement of the celebrated Bloch-Nordsieck
model where all the spin-related effects are ignored which makes
this model exactly soluble but restricts its applicability to
the infrared (IR) regime $|p^2-m^2|\ll m^2$ near the fermion's mass shell.

We emphasize that the IR regime can only exist if the fermions are massive,
while in the massless case the entire region below the upper cutoff $\Lambda$ 
(which is set by the conditions of the applicability of the effective 
$QED$-like description itself) falls into the opposite, ultraviolet (UV), 
regime which, in the case of a finite fermion mass, is defined
as $|p^2-m^2|\gg m^2$.

The rest of the paper is organized as follows.
We first describe the Schwinger's functional technique
and investigate both the IR and UV asymptotics of the ordinary (gauge-dependent) 
fermion Green function in the general D-dimensional case. 
Then, after having compared our general formulas with the well known 3D results as well as 
with the partially known 2D ones, we proceed with the gauge-invariant
fermion amplitude proposed in Ref.\cite{Wen} and ascertain its true behavior. 
We conclude our analysis with a discussion of the alternatives to the previously suggested  
form of the physical electron propagator as well as to the fits to 
the ARPES data \cite{ARPES} exploiting the $QED_{2+1}$-related scenarios.

\section{Functional integral representation of fermion amplitudes} 
The conventional fermion Green function is given by the (properly normalized)
functional integral over the fermion and gauge field configurations
\be
G(x, y)= <\Psi(x) {\overline \Psi}(y)>=
\int D[{\overline \Psi}]D[{\Psi}]D[{\bf A}] \Psi(x){\overline \Psi}(y)
\exp(iS[\Psi,{\overline \Psi},{\bf A}])
\ee
Upon integrating the fermions out, one arrives at the expression 
\be
G(x, y)= \int D[{\bf A}]G(x, y|{\bf A})\exp(iS_{eff}[{\bf A}])
\ee
where the effective action of the gauge field includes the fermion polarization   
$$
S_{eff}[{\bf A}]=
{1\over 2g^2}\int d{\bf x}( \partial_\mu A_\nu - \partial_\nu A_\mu )^{2} +
\ln { {\it det}[i{\hat \partial}+{\hat A}-m]\over
{ {\it det}[i{\hat \partial}-m] } }
$$
\be
={1\over 2}\int d{\bf x}\int d{\bf y}A_\mu (x) {\cal D}^{-1}_{\mu\nu}(x-y)A_\nu (y)+\dots
\ee
By neglecting all but the gaussian term in (7) one
excludes from consideration any processes of "light-light scattering" and alike.
Thus far, none of the beforementioned effective $QED$-like 
descriptions of the condensed matter systems has gone anywhere
beyond this common approximation.

Nonetheless, the gauge field is not completely quenched, as 
one still accounts for the quadratic polarization $\Pi(q)$, resulting 
in  the gauge field propagator
which, in the covariant $\lambda$-gauge, assumes the form 
\be
{\cal D}_{\mu\nu}(q)={g^2\over q^2+\Pi(q)}(\delta_{\mu\nu}+(\lambda-1){q_\mu q_\nu\over q^2})
\ee
In turn, the fermion Green function $G(x, y|{\bf A})$ computed 
for a given gauge field configuration obeys the equation
\be
[i{\hat \partial}+{\hat A}(x)-m] G(x, y |{\bf A}) = \delta({\bf x}-{\bf y})
\ee
Its formal solution can be written in the form of a quantum mechanical (i.e., $single-particle$)
path integral \cite{Lebedev} 
\bea
G(x, y| {\bf A})=-i\int^\infty_0 ds e^{is(-m^2+i\delta)}
[i{\hat \partial}+{\hat A}(x)+m]
\int D[{\bf {a}}]\delta({\bf x}-{\bf y}-2\int^s_0{\bf {a}}(\tau_2)d\tau_2)
\nonumber\\
\exp[-i\int^s_0d\tau ({\bf {a}}^2(\tau)-(2a_\mu(\tau)+
\sigma_{\mu\nu}i\partial_\nu)
A_\mu ({\bf x}-2\int^s_\tau {\bf {a}}(\tau_1)d\tau_1 ) ) ]
\eea
where $\sigma_{\mu\nu}=[\gamma_\mu,\gamma_\nu]/2$ and $\delta\to 0^{+}$.
The integral over the fermion's momentum ${\bf a}(s)$ 
as a function of the proper time $s$ parameterizing its space-time trajectory is  
normalized in such a way that 
$$
\int D[{\bf a}]\exp(-i\int^s_0{\bf a}^2(\tau))d\tau=1
$$
Next, we perform functional averaging over different 
gauge field configurations with the use of Eq.(7),
then Fourier transform Eq.(10) to the momentum representation,
and finally switch to the integration over the fluctuating part of the total fermion's 
momentum ${\bf v}(s)={\bf {a}}(s)-{\bf p}$, thus obtaining
\be
G(p)=-i\int^\infty_0 ds e^{is(p^2-m^2+i\delta)}
\int D[{\bf v}]\exp(-i\int^s_0{\bf v}^2(\tau)d\tau)
[{\hat p}+m+M(s|{\bf v})]\exp(i\Phi(s|{\bf v}))
\ee
In this expression, the terms which are odd in ${\bf A}(z)$ 
contribute to the gauge invariant (see below) part of the mass operator 
\be
M(s|{\bf v})=\int {d{\bf q}\over (2\pi)^{D+1}} {\cal D}_{\mu\nu}(q)\int^s_0 d\tau 
\gamma_\mu (2v_\nu(\tau)+2p_\nu-\sigma_{\nu\lambda}q_\lambda)
e^{2i{\bf p}{\bf q}(s-\tau)+2i\int^s_\tau {\bf q}{\bf v}(\tau_1)d\tau_1}
\ee
while the even ones stem from the 
exponential of the (gauge-dependent) "phase factor" 
\bea
\Phi(s|{\bf v})=
\int {d{\bf q}\over (2\pi)^{D+1}} {\cal D}_{\mu\nu}(q)\int^s_0 d\tau_1 \int^{\tau_1}_0 d\tau_2
(2v_\mu(\tau_1)+2p_\mu+\sigma_{\mu\alpha}q_\alpha)
(2v_\nu(\tau_2)+2p_\nu-\sigma_{\nu\beta}q_\beta)
\nonumber\\
e^{2i{\bf p}{\bf q}(\tau_1-\tau_2)+2i\int^{\tau_1}_{\tau_2} 
{\bf q}{\bf v}(\tau_3)d\tau_3}
\eea
In the above expressions, the integrations over the proper time parameters $\tau_i$
are ordered according to the order of their appearance in the products of the 
non-commutative factors $(2v_\mu(\tau_i)+2p_\mu\pm\sigma_{\mu\nu}q_\nu)$.

\section{Infrared behavior}
By using Eqs.(11-13) one can readily determine the IR behavior
of the fermion Green function. With its momentum satisfying the condition $|p^2-m^2|\ll m^2$  
a fermion behaves as a heavy particle
whose velocity remains essentially unchanged atfer emitting and absorbing an arbitrary 
number of the gauge field quanta.
Therefore, the Green function receives its main contribution 
from the fermion trajectories close to the straight-line path
(which only coincides with the semiclassical trajectory in the case 
of a time-like separation between the ending points $(x-y)^2>0$).

This allows one to neglect the fluctuations of the total fermion's momentum
with respect to its average value $\bf p$, 
in which case the mass operator introduces only a small correction
\be
M_{IR}(s|{\bf v})=i\int {d{\bf q}\over (2\pi)^{D+1}}{\cal D}_{\mu\nu}(q)\gamma_\mu p_\nu
{1-e^{2i{\bf q}{\bf p}s}\over {\bf q}{\bf p}}={\hat p}O({1\over sp^2})\sim {\hat p}
{|p^2-m^2|\over m^2}\ll {\hat p}
\ee
In deriving (14) we took into account that a characteristic value of the parameter
$s\sim |p^2-m^2|^{-1}$ is determined by Eq.(11) and the fact that the integral 
(14) receives its main contribution from small transferred momenta
$
q\lesssim {1/sp}\sim {|p^2-m^2|/p}\ll p
$.
 
In contrast, the integrals over $\tau_i$ in the 
gauge-dependent IR phase factor are formally divergent. 
They must be tackled by first computing the momentum integral 
and then applying the so-called "ribbon" regularization  \cite{Stefanis} 
${\bf p}(\tau_1-\tau_2)\to {\bf p}(\tau_1-\tau_2)
+{\bf l}$ with $({\bf p}{\bf l})=0$ and $|{\bf l}|=1/\Lambda$ which yields the expression 
$$
\Phi_{IR}(s)=
4\int {d{\bf q}\over (2\pi)^{D+1}} {\cal D}_{\mu\nu}(q)
\int^s_0 d\tau_1 \int^{\tau_1}_0 d\tau_2
p_\mu p_\nu e^{2i{\bf p}{\bf q}(\tau_1-\tau_2)}
$$
\bea
=i{g^2}I_D\int^s_0 d\tau_1 \int^{\tau_1}_0 d\tau_2
[(D-2+\lambda){{p}^2\over |{\bf p}(\tau_1-\tau_2)+{\bf l}|^2}-
2(\lambda-1){p^4(\tau_1-\tau_2)^2\over |{\bf p}(\tau_1-\tau_2)+{\bf l}|^4}]
\nonumber\\
=ig^2I_D[(D-2+\lambda)({\pi\over 2}(sp\Lambda)-\ln(sp\Lambda))-
2(\lambda-1)({\pi\over 4}(sp\Lambda)-\ln(sp\Lambda))]
\eea
In the massive case, the linear divergence of $\Phi(s)$ would be  
routinely attributed to the renormalization of the bare mass $m\to m+O(\Lambda)$.
After having separated this linear divergence,
we observe that the subleading logarithmic terms 
conspire to give rise to the non-perturbative formula 
\be
G_{IR}(p)=-i({\hat p}+m)
\int^\infty_0 ds e^{is(p^2-m^2+i\delta)}(sp\Lambda)^{-\eta_{IR}/2}
\sim {{\hat p}+m\over
(p^2-m^2+i\delta)^{1-\eta_{IR}/2}}
\ee
which, near the mass shell,
exhibits the anticipated algebraic behavior (3) with the IR anomalous dimension 
\be
\eta_{IR}=2g^2I_D(\lambda-D)
\ee
where 
\be
I_D=[2^{D}\pi^{(D+1)/2}\Gamma((D+1)/2)]^{-1}
\ee
Thus, in the 3D case of the conventional weakly coupled
$QED_{3+1}$ we recover the well-known IR exponent 
(see, e.g., \cite{Landau})
\be
\eta^{3D}_{IR}={e^2\over 4\pi^2}
(\lambda-3)
\ee
which vanishes in the so-called Yennie's gauge $\lambda=3$ ($\eta_{IR}$ is also known
to be zero in some non-covariant gauge, such as the Coulomb gauge ${\vec q}{\vec A}=0$).

In the (parity-even) 2D case which is of a particular interest in view 
of its condensed matter-related
applications \cite{Lee,Wen,Franz,Ye,Herbut,graphite,DVK},
the weak coupling regime turns out to be intrinsically unstable against 
the effects of the fermion polarization. In fact,
for $q\lesssim Ng^2$ the gauge propagator is totally dominated  by the fermion polarization
which, for $N\gg 1$, is given by the one-loop term 
\be
\Pi(q)={Ng^2\over 8}{\sqrt {-q^2}}
\ee
and the gauge field propagator reads as 
\be
{\cal D}^{2D}_{\mu\nu}(q)={8\over N{\sqrt {-q^2}}}
(\delta_{\mu\nu}+(\lambda-1){q_\mu q\nu\over q^2})
\ee 
Instead of the bare coupling $g$, it is $1/N$
that now becomes a parameter of the perturbative expansion. 
We note that above the momentum scale $Ng^2$ no further logarithmic corrections 
are generated, so that the latter is now playing the role of the UV cutoff.
Nonetheless, for the sake of uniformity of our presentation,
in the following discussion
 we will continue using 
the notation $\Lambda$ and the label UV
for the range of momenta $m\ll q\lesssim \Lambda=Ng^2$.

It is also worth mentioning that,
owing to the parity conserving structure of the reducible four-fermion representation,
the radiative corrections generate no Chern-Simons terms.

Using (21) we obtain a coupling-independent anomalous exponent   
\be
\eta^{2D}_{IR}={8\over \pi^2N}(\lambda-2),
\ee 
thus discovering the 2D analogue ($\lambda=2$) of the 3D Yennie's gauge. 

Notably, the IR wave function renormalization assumes the anticipated power-law form,
in full accord with the physical origin of the IR singularity. The latter is known
to stem from the processes involving independent emission and absorption of
an arbitrary large number of soft gauge quanta. Due to their uncorrelated 
nature, these multiple "bremsstrahlung" events obey a Poisson distribution formula,
hence the appearance of the factorials in the statistical weights, resulting in 
the natural exponentiation of the lowest order ($\sim g^2\ln\Lambda$) correction.

\section{Ultraviolet behavior}
The Schwinger's functional technique is also capable of exploring the UV regime  
($|p^2-m^2|\gg m^2$) which is the only regime of interest present 
in the massless case. Despite the fact that the procedure
is straightforward, there seems to have been no
such a systematic attempt made in the past.  

Technically, the UV behavior is more difficult to analyze, because
the path integral (11) is no longer saturated by the trajectories close to the semiclassical
straight line. In fact, the relevant paths can strongly deviate 
from the straight-line one, for they suffer no exponential suppression,
unlike in the IR regime.

Despite the fact that 
the functional integration over ${\bf v}(s)$ can no longer be carried out exactly,
one can instead resort to the formula
\be
\int D[{\bf v}]e^{-i\int {\bf v}^2d\tau +F[{\bf v}]}
=
e^{<F>}\int D[{\bf v}] e^{-i\int {\bf v}^2d\tau} \sum^\infty_{n=0}
{(F[{\bf v}]-<F>)^2\over n{!}}
\ee
where
$
<F>=\int D[{\bf v}]e^{-i\int {\bf v}^2d\tau} F[{\bf v}]
$.

Eq.(23) has been extensively used, e.g., in implementing the 
Feynman's variational principle in the polaron and related problems.
Expanding (11) to the first order in ${\cal D}_{\mu\nu}(q)$ we obtain
\be
\delta_1 G_{UV}(p)=
-i\int^\infty_0 ds e^{is(p^2+i\delta)}
[<M_{UV}(s)>+i{\hat p}<\Phi_{UV}(s)>]
\ee
The functionally averaged
mass operator (12) is now determined by the transferred momenta $q\gg p\sim 1/{\sqrt s}$
and it needs to be computed only to the first order in $\bf p$ 
\be
<M_{UV}(s)>
=i\int {d{\bf q}\over (2\pi)^{D+1}}
{\cal D}_{\mu\nu}(q){1-e^{is(q^2+2{\bf q}{\bf p})}\over
q^2+2{\bf q}{\bf p}}\gamma_\mu(q_\nu+2p_\nu-\sigma_{\mu\lambda}q_\lambda)
=2g^2{\hat p}I_D{D\over D+1}\ln(s\Lambda^2)+\dots
\ee
Notably, Eq.(25) is independent of the gauge parameter. 
In contrast, the averaged phase factor (13) 
which can be calculated in the ${\bf p}\to 0$ limit  
$$
<\Phi_{UV}(s)>=\int {d{\bf q}\over (2\pi)^{D+1}}{{\cal D}_{\mu\nu}(q)}
$$
\be
[{1-e^{is(q^2+2{\bf q}{\bf p})}+is(q^2+2{\bf q}{\bf p})
\over (q^2+2{\bf q}{\bf p})^2}
(q_\mu+2p_\mu+\sigma_{\mu\alpha}q_\alpha)
(q_\nu+2p_\nu-\sigma_{\nu\beta}q_\beta)
-{i}s\delta_{\mu\nu}]
={i\over 2}g^2I_D(D+\lambda)\ln(s\Lambda^2)+\dots,
\ee
does manifest a dependence on the gauge parameter.
Combining (25) and (26) together, we obtain the total correction to the Green function  
$$
\delta_1 G_{UV}(p)=\int^\infty_0 ds e^{is(p^2+i\delta)}\int {d{\bf q}\over (2\pi)^{D+1}} 
{{\cal D}_{\mu\nu}(q)}[{1-e^{is(q^2+2{\bf q}{\bf p})}\over {q^2+2{\bf q}{\bf p}}}
[\gamma_\mu (q_\nu+2p_\nu-\sigma_{\nu\lambda}q_\lambda)
$$
$$
+
{\hat p}{1-e^{is(q^2+2{\bf q}{\bf p})}+is(q^2+2{\bf q}{\bf p})\over (q^2+2{\bf q}{\bf p})^2}
(q_\mu+2p_\mu+\sigma_{\mu\alpha}q_\alpha)
(q_\nu+2p_\nu-\sigma_{\nu\beta}q_\beta)
-
i{\hat p}s\delta_{\mu\nu}]
$$
\be
={g^2\over 2}{{\hat p}\over p^2}I_D({D(3-D)\over D+1}-\lambda)
\ln({\Lambda^2\over p^2})+\dots 
\ee
By using the identity 
$$
{\hat p}\gamma_\mu({\hat p}+{\hat q})\gamma_\nu{\hat p}=
{\hat p}(q_\mu+2p_\mu+\sigma_{\mu\alpha}q_\alpha)
(q_\nu+2p_\nu-\sigma_{\nu\beta}q_\beta)
-\gamma_\mu p^2(q_\nu+2p_\nu-\sigma_{\nu\lambda}q_\lambda)-\delta_{\mu\nu}{\hat p}(p+q)^2
$$
and integrating in (27) over the proper time $s$ prior to the 
momentum integration one can also check that the correction given by Eq.(27) exactly reproduces
the one-loop result of the conventional diagrammatic expansion 
\be
\delta_1G_{UV}(p)=-i\int {d{\bf q}\over (2\pi)^{D+1}} 
{{\cal D}_{\mu\nu}(q)\over p^4(p+q)^2}
{\hat p}\gamma_\mu({\hat p}+{\hat q})\gamma_\nu{\hat p}
\ee
Instead of expanding Eq.(11) to higher orders in ${\cal D}_{\mu\nu}(q)$ one can perform
summation of the leading $(g^{2}\ln\Lambda)^n$ terms
by virtue of the standard renormalization group equation which, 
reflects the scaling properties of a generic two-point amplitude 
(gauge invariant and non-invariant alike) 
under the change of the upper cutoff \cite{Landau} 
\be
[\Lambda{\partial\over \partial\Lambda}-\beta({\tilde g}){\partial\over\partial {\tilde g}}+
\eta({\tilde g})]{\hat p}G_{UV}(p; \Lambda; {\tilde g}) =0 
\ee
where the leading order dependence of the anomalous dimension of the fermion Green function 
on the renormalized coupling strength 
$\tilde g$ is given by the explicit form of the first order correction (27)  
\be
\eta(g)=-\Lambda{\partial\over \partial\Lambda}{\hat p}\delta_1G_{UV}(p; \Lambda; g)|_{p=\Lambda}
\ee
while $\beta(g)=\Lambda{\partial {\tilde g}/\partial\Lambda}|_{p=\Lambda}=0$, 
and, therefore, the coupling strength
retains it bare value ${\tilde g}=g$,
for as long as the dynamics of the gauge field is considered quenched. 

The solution of Eq.(29) suggests that the first logarithmic correction (27) merely gets 
exponentiated, thus yielding the algebraic behavior controlled by the UV exponent
\be
\eta_{UV}={g^2I_D}(\lambda+{D(D-3)\over D+1})
\ee
Further corrections to Eq.(31) require one not only to extract  
the subleading corrections of order $g^{2n}\ln\Lambda$ from the $n^{th}$-order
terms in the expansion of Eq.(11) in powers of ${\bf v}(s)$ 
and account for the improved fermion polarization $\Pi(q)$ 
but also to proceed
beyond the quenched approximation (7) for the effective action of the gauge field.
 
In the weakly-coupled 3D case Eq.(31)
reproduces the well known result \cite{Landau} 
\be
\eta^{3D}_{UV}={e^2\over 8\pi^2}\lambda
\ee
while in the $2D$ case it yields the coupling-independent UV exponent 
\be
\eta^{2D}_{UV}={4\over 3\pi^2N}(3\lambda-2), 
\ee
in agreement with the result obtained in \cite{Nash}.

\section{Gauge invariant fermion amplitude}
After having tested our formalism against the known examples, we turn to 
the proposed candidate for the physical electron propagator which is given by   
Eq.(2) with the straight-line contour $\Gamma$
\be
G^{|}_{inv}(x, y)=\int D[{\bf A}]G(x, y|A)\exp(i\int^x_y dz^\mu A_\mu(z))\exp(iS_{eff}[A])
\ee
Proceeding by analogy with the derivation presented in Section II,
 one readily obtains Eq.(11) where
Eqs.(12) and (13) are replaced, respectively, with
\bea
M_{inv}(s|{\bf v})=
\int{d{\bf q}\over (2\pi)^{D+1}} {\cal D}_{\mu\nu}(q)
\int^s_0 d\tau \gamma_\mu
[(2v_\nu(\tau)+2p_\nu-\sigma_{\nu\lambda}q_\lambda)
e^{2i{\bf p}{\bf q}(s-\tau)+2i\int^s_\tau {\bf q}{\bf v}(\tau_1)d\tau_1}
\nonumber\\
-
(2v_\nu(\tau_2)+2p_\nu-\sigma_{\nu\lambda}q_\lambda)
e^{2i{\bf p}{\bf q}\tau + 2i{\tau/s}\int^s_0 {\bf q}{\bf v}(\tau_1)d\tau_1}
\int^s_0 {d\tau_2\over s}]
\eea
and
$$
\Phi_{inv}(s|{\bf v})=\int 
{d{\bf q}\over (2\pi)^{D+1}}
{\cal D}_{\mu\nu}(q) 
\int^s_0 d\tau_1\int^{\tau_1}_0 d\tau_2 
(2v_\mu(\tau_1)+2p_\mu+\sigma_{\mu\alpha}q_\alpha)
(2v_\nu(\tau_2)+2p_\nu-\sigma_{\nu\beta}q_\beta)
$$
$$
[e^{2i{\bf p}{\bf q}(\tau_1-\tau_2)+2i\int^{\tau_1}_{\tau_2}{\bf q}{\bf v}(\tau_3)d\tau_3}
+2\int^s_0 {d\tau_3\over s} \int^{\tau_3}_0 {d\tau_4\over s} 
e^{ 2i{\bf p}{\bf q}(\tau_3-\tau_4)+2i(\tau_3-\tau_4)/s 
\int^{s}_{0}{\bf q}{\bf v}(\tau_5)d\tau_5}
$$
\be
-
2\int^s_0 {d\tau_3\over s} 
e^{2i{\bf p}{\bf q}(s-\tau_1)+2i\int^{s}_{\tau_1}{\bf q}{\bf v}(\tau_4)d\tau_4
-2i{\bf p}{\bf q}\tau_3-2i{\tau_3/s}\int^s_0{\bf q}{\bf v}(\tau_5)d\tau_5}]
\ee
In the IR regime the path integration can still be carried out
exactly by simply neglecting ${\bf v}(s)$ with respect to the average fermion's
momentum $\bf p$.
In the same approximation as that used in Section III (which is only
justified in the vicinity of the mass shell, provided that $m\neq 0$), one 
readily obtains 
\bea
M_{inv,IR}(s)=2
\int {d{\bf q}\over (2\pi)^{D+1}}
{\cal D}_{\mu\nu}(q)\int^s_0 d\tau 
\gamma_\mu p_\nu [e^{2i{\bf p}{\bf q}(s-\tau)}
-e^{2i{\bf p}{\bf q}\tau}]=0
\eea
and
$$
\Phi_{inv,IR}(s)=
4\int {d{\bf q}\over (2\pi)^{D+1}}
{\cal D}_{\mu\nu}(q)p_\mu p_\nu
\int^s_0 d\tau_1\int^{\tau_1}_0 d\tau_2
$$
\be 
[e^{2i{\bf p}{\bf q}(\tau_1-\tau_2)}+
{2\over s^2}\int^s_0 d\tau_3 \int^{\tau_3}_0 d\tau_4 
e^{2i{\bf p}{\bf q}(\tau_3-\tau_4)}
-{2\over s}\int^s_0 d\tau_3 
e^{2i{\bf p}{\bf q}(s-\tau_1-\tau_3)}]=0
\ee
Thus, as first pointed out by the authors of Refs.\cite{Lebedev},
in the IR regime the gauge-invariant propagator
(34) retains a simple pole 
\be
G^{|}_{inv, IR}(p)\approx {{\hat p}+m\over p^2-m^2+i\delta}
\ee
hence, $\eta_{inv,IR}=0$.

By comparing (39) and (17) one can also deduce the IR anomalous dimension of the exponential
factor $\exp(i\int dx_\mu A_\mu)$ itself
\be
\eta_{exp,IR}={2g^2I_D}(D-\lambda)
\ee
which of course vanishes in the Yennie's gauge.

Next, going over to the UV regime and expanding Eqs.(35) and (36)
to the first order in ${\cal D}_{\mu\nu}(q)$ we arrive at
Eq.(24) where
the functional average of the gauge-dependent phase factor 
$$
<\Phi_{inv,UV}(s)>=\int{d{\bf q}\over (2\pi)^{D+1}}
{g^2\over q^2+\Pi(q)}
$$
\be
[{1-e^{is(q^2+2{\bf q}{\bf p})}+is(q^2+2{\bf q}{\bf p})\over (q^2+2{\bf q}{\bf p})^2}
({p^2q^4\over ({\bf q}{\bf p})^2}-{(\sigma_{\mu\nu}q_\nu)^2}
-q^2)-is]=O(p^2s)\lesssim 1
\ee
now exhibits neither linear, nor logarithmic divergence as a function of $s$, 
unlike in the case of the non-invariant amplitude (see Eq.(26)). 
In turn, the value of the mass operator 
\bea
<M_{inv,UV}(s)>=2i\int {d{\bf q}\over (2\pi)^{D+1}}{g^2\over q^2+\Pi(q)}
{1-e^{is(q^2+2{\bf q}{\bf p})}\over q^2+2{\bf q}{\bf p}}
[{\hat p}-{\hat q}{p^2\over {\bf q}{\bf p}}-{\hat q}{p_\mu\sigma_{\mu\nu}q_\nu\over q^2}
+{{\bf q}{\bf p}\over q^2}{\gamma_\mu\sigma_{\mu\nu}q_\nu}]
\nonumber\\
=2g^2{\hat p}I_D {D\over D+1}\ln(s\Lambda^2)+\dots 
\eea
appears to coincide with Eq.(25). 
Thus, it is Eq.(42) that 
solely determines the correction to the gauge-invariant Green function  
\be
\delta_1 G_{inv, UV}(p)=
-i\int^\infty_0 ds e^{is(p^2+i\delta)}
[<M_{inv,UV}(s)>+i{\hat p}<\Phi_{inv,UV}(s)>]=
2{g^2}{{\hat p}\over p^2}I_D{D\over D+1}\ln({\Lambda^2\over p^2})
\ee

The same result can be obtained by working in the axial
gauge ${\bf n}{\bf A}=0$ defined by the vector ${\bf n}=({\bf x}-{\bf y})/|x-y|$. 
In this gauge, the exponential factor in (34) is identically equal to unity,
and the first order correction is given by Eq.(28) where one has to
use the gauge field propagator
\be
{\cal D}^{ax}_{\mu\nu}(q)={g^2\over q^2+\Pi(q)}
[\delta_{\mu\nu}+n^2
{q_\mu q_\nu\over ({\bf n}{\bf q})^2}-
{n_\mu q_\nu+q_\mu n_\nu\over ({\bf n}{\bf q})}]
\ee
Notably, the result (43) obtained with the use of Eq.(28) 
is independent of the direction of the vector $\bf n$,
for all the terms proportional to ${\hat n}({\bf n}{\bf p})$ cancel out
and only those proportional to $\hat p$ remain in the final expression.

It is worth mentioning that the integrals 
in Eqs.(41,42) as well as in Eq.(28) with 
the gauge propagator (44) are all plagued with the spurious poles, such as
 $1/({\bf q}{\bf p})^{1,2}$.
We handle these singular denominators by resorting to the exponential integral representation: 
$1/({\bf q}{\bf n})=-i\int^\infty_0 ds \exp(is({\bf q}{\bf n}+i\delta))$.
Then, after having performed the Lorentz-invariant momentum 
integration, we carry out the remaining 
integrals over the auxiliary parameter $s$ with the use of the "ribbon" regularization
\cite{Stefanis}. This procedure yields the following logarithmic integrals
appearing in our calculation
$$
\int {{d{\bf q}\over (2\pi)^{D+1}}}{q_\mu\over q^{D-1}(p+q)^2({\bf q}{\bf n})}
={iI_D\over 2}{n_\mu\over n^2}\ln ({\Lambda^2\over p^2}),
$$ $$
\int {d{\bf q}\over (2\pi)^{D+1}}
{q_\mu q_\nu\over q^{D-1}(p+q)^2({\bf q}{\bf n})^2}=
{iI_D\over 2}{2n_\mu n_\nu - \delta_{\mu\nu}{n}^2\over n^4}\ln ({\Lambda^2\over p^2}),
$$
and 
$$
\int {d{\bf q}\over (2\pi)^{D+1}}
{q_\mu q_\nu q_\lambda\over q^{D+1}(p+q)^2({\bf q}{\bf n})}=
{iI_D\over 2(D+1)}
({n_\mu \delta_{\nu\lambda}+ n_\nu \delta_{\mu\lambda}+n_\lambda \delta_{\mu\nu}\over n^2}
 -2{n_\mu n_\nu n_\lambda\over n^4})\ln ({\Lambda^2\over p^2})
$$
One can check that the above expressions are fully
consistent 
with the standard "principal value" prescription for the spurious poles, whose  
advanced form is known in the field-theoretical 
literature as the Leibbrandt-Mandelstam rule (see \cite{Lavelle} and references therein).

Finally, by invoking the renormalization group equation (29) we find that 
the logarithmic correction (43) tends to exponentiate, 
thereby resulting in the new UV anomalous dimension
\be
\eta_{inv,UV}=-4g^2I_D {D\over D+1},
\ee
which appears to be $negative$.

Surprisingly enough, we were unable to find in the literature any result 
pertaining to the weakly-coupled
3D abelian gauge theory (e.g., conventional $QED_{3+1}$), in which case Eq.(45) yields 
\be
\eta^{3D}_{inv,UV}=-{3g^2\over 8\pi^2}
\ee
Nevertheless, we did find some comfort in comparing (46) with the exponent 
which had been previously 
found to control the power-law UV behavior of the non-abelian  analogue of Eq.(34) in the 
$SU(3)$-symmetrical case \cite{Dorn} 
\be
\eta^{3D, SU(3)}_{inv,UV}=-{g^2\over 2\pi^2}
\ee
By construction, Eq.(47) is proportional to the quadratic Casimir operator in  
the fundamental representation of the color group which, in the case of $SU(N)$, equals  
\be
c_F={1\over N}\sum_{a=1}^{N^2-1}tr( T^aT^a)={N^2-1\over 2N}
\ee
Evaluating (48) for $SU(3)$ we obtain $c_F^{SU(3)}=4/3$ and,   
upon separating this factor out, recover the result (46) pertinent to the abelian case
(with the electric charge $e$ substituted for $g$).
 
Likewise, by using Eq.(21) we obtain the anomalous exponent which 
controls the gauge-invariant propagator in $QED_{2+1}$   
\be
\eta^{2D}_{inv,UV}=-{32\over 3\pi^2N}
\ee
which is negative, contrary to the result of Ref.\cite{Wen}
and in agreement with the sign (albeit not the magnitude) 
of the exponent quoted without derivation in Ref.\cite{Franz}.
However, it remains to be seen whether the exponentiation of $M^{UV}_{inv}(s)$ as well
as vanishing of $\Phi^{UV}_{inv}(s)$ still hold beyond the leading $1/N$ order.

Lastly, by comparing Eqs.(31) and (45) one can also deduce the UV anomalous dimension of 
the exponential factor $\exp(i\int A_\mu dz_\mu)$ 
\be
\eta_{exp,UV}=-{g^2I_D}(D+\lambda)
\ee
Interestingly enough, for $\lambda=-D$ this exponent equals zero, 
and the UV anomalous dimension of
the non-invariant propagator coincides with (45), in agreement with the 
observation made in the 3D non-abelian case \cite{Dorn}.

\section{Discussion}
Our calculation demonstrates that in the massless case  
the gauge invariant Green function (34) appears to decay $slower$ than the bare one,
in a marked contrast with the previously conjectured Luttinger-like behavior.
In this concluding Section, we make an attempt to rationalize these findings,
although we refrain from making any final judgement on their physical implications. 

Albeit somewhat counterintuitive, the found UV behavior is not totally incomprehensible.
In fact, the generic 
behavior of an invariant fermion amplitude is manifested by the asymptotic formula 
\be
G_{inv}^{\Gamma}(x) \sim \exp(-C|x|\Lambda+\eta \ln(|x|\Lambda))
\ee
where $C>0$, and the expression (51) decays with $|x|$ exponentially, regardless of the sign
of $\eta$, because the logarithmic term in the exponent is subleading to the 
linear one. However, in a renormalizable gauge theory where
the gauge invariance is reinforced throughout the whole process of renormalization, 
the latter would be routinely cancelled out by counterterms, which leaves behind the 
logarithmic term of (potentially) either sign.

This situation would change, however, should one choose to relaxe the condition
of renormalizability at the expense of the gauge invariance,
since the radiative corrections to 
the action (1) generically produce a finite mass of the vector field $A_\mu$.  
Loosely speaking, 
the situation would then resemble that in the Schwinger's $QED_{1+1}$ where the gauge field
acquires a mass $M\sim g$, and the analogue of Eq.(34) behaves as 
\be
G^{|}_{inv}(x)\sim \exp(-{1\over 2}[\ln(Mx)+K_0(Mx)]-Mx)
\ee
It is worth noting that, should one decide to intentionally
disregard the exponential factor $e^{-Mx}$, Eq.(52) would appear to 
exhibit a power-law decay $\sim 1/{\sqrt x}$ at $x\gg 1/M$, thus 
siggesting $\eta^{1D}_{inv}=-1/2$.

We mention, in passing, that the exponential, rather than a power-law, behavior
has also been found in the
problem of Dirac fermions in 
the presence of a static random vector potential (${\bf A}(x)=(0,{\vec a}({\vec r}))$) 
which allows for an asymptotically exact
solution in the ballistic regime of large fermion energies \cite{us}.
   
Conceivably, in some of  
the abovementioned physical applications of $QED_{2+1}$ with $N=2$ the problem
of the slow space-time decay of the gauge invariant amplitude (34) can be thwarted 
by a spontaneous development of a finite fermion mass,
in which case the behavior of $G^{|}_{inv}(x)$ at large $x$ will be governed by the (free)
IR asymptotic (39) instead of the UV one. However, the intrinsic
propensity of the 2D Dirac fermions in $QED_{2+1}$ towards generating a finite mass 
(usually referred to as the phenomenon of chiral symmetry breaking)
is believed to occur only at sufficiently small $N<N_c$ \cite{Appelquist}.  
While in the case of the Lorentz-invariant
action (1) the critical number of flavors $N_c$ was found to be as low as $3/2$ 
\cite{Appelquist}, the Lorentz-(or even rotationally-) 
non-invariant generalizations of the action (1) are still awaiting to be fully explored. 

To this end, the authors of Refs.\cite{Herbut} conjectured that 
the critical value $N_c$ in the $QED$-like description of the quantum disordered
planar $d$-wave superconductor may become greater than two
due to the lack of rotational invariance.
On the other hand, in the finite-temperature counterpart of the 
2D chiral symmetry breaking transition in the (spatially) rotationally-invariant  
effective theory of a single layer of graphite 
$N_c$ was found to be further reduced as compared to the Lorentz-invariant case \cite{DVK}.

However, should one insist on   
maintaining both the gauge and Lorentz invariances of the renormalized
gauge field action, the problem of the slow spatial decay of the 
alleged physical electron propagator (34) 
associated with its negative UV anomalous dimension (45) could not be resolved
without re-examining the "minimal" form of this Green function. In fact, the task of 
constructing the proper gauge transformation which converts the auxiliary
Dirac fermions into the physical electrons may not be limited
to a particular choice of the contour $\Gamma$ in Eq.(2) but may also 
require one to modify the phase factor itself. 

It is worth noting that in the previous calculations of the "zero-bias anomaly"
in the tunneling density of states in the compressible Quantum Hall effect \cite{Kim},
the construction of the electron Green function, albeit seemingly given by 
the same Eq.(34) with the contour $\Gamma$ now chosen along the temporal axis,
was, in fact, more involved. 
Indeed, in the semiclassical approximation employed in \cite{Kim}, 
the gauge field dependence of the exponential factor $\exp(i\int dz_\mu A_\mu)$
would have been exactly compensated by that of 
the non-gauge invariant Green function  $G(t,{\vec 0}|{\bf A})$, thus making 
the functional average of the product of the two behave essentially as in the absence
of any gauge coupling. 

Nevertheless, the electron density of states computed in \cite{Kim}
appeared to be strongly affected by the 
Chern-Simons gauge fluctuations which can be traced back to the fact
that, in addition to the abovementioned factors, the electron Green function 
happened to contain yet another factor, the exponent of the saddle-point value of the   
effective action of the Chern-Simons gauge field. It was, in fact, this factor which 
was solely responsible for the strong suppression of the tunneling density of states,
consistent with the physical interpretation of the Chern-Simons field
as representing the effect of the Coulomb coupling in the presence of strong magnetic field.
In light of the fact that in the problem at hand the time reversal symmetry
remains unbroken, no such an additional factor can be readily  
incorporated into the naive form of the electron propagator (34).

In order to further elaborate on this point, we mention  
yet another example demonstrating the sensitivity 
of a generic gauge invariant amplitude to the details of its construction.
To this end, we recall the original Dirac's idea of explicitly constructing a 
"dressed charge" corresponding to a physical electron by means of the gauge 
transformation 
\be
\Psi_{phys}(x)=\exp(i\int d{\bf y}\chi_\mu(x-y)A_\mu(y))\Psi(x)
\ee
where the vector function 
$\chi_\mu(x)$ obeys the equation $\partial_\mu\chi_\mu(x)=\delta(x)$.
In the time-independent Shrodinger operator representation,
the originally proposed transformation from the bare fermions to the physical
electrons was implemented as a space-like Dirac string between the location of the  
fermion and an infinitely remote point   
$$
\chi_0=0,~~~~~\chi_i=<{\vec x}|{1\over {\bf \nabla}^2}\partial_i|{\vec y}>
$$
The Fourier transform of the electron propagator 
$<\Psi_{phys}(x){\overline \Psi}_{phys}(y)>$ is IR finite (see Eq.(39)) and undergoes 
multiplicative UV renormalization at a single point $p_\mu=(m, {\vec 0})$ 
on the mass shell corresponding to a static charge \cite{Lavelle}, in agreement
with the general expectation that 
the absence of any singularity other than a simple pole is characteristic 
of the propagator of an exact eigenstate with the quantum numbers
of electron. 

It was shown in \cite{Lavelle} that in the case
of a dressed charge moving with a finite velocity $\vec u$
the above phase factor needs to be further modified 
\be
\Psi_{phys}(x|{\vec u})=
\exp(i\gamma\int d^{D-1}{\vec y}_{\perp}dy_{\parallel}
<{\vec x}|{1\over {\bf \nabla}^2}|{\vec y}>
[\gamma^{-2}\partial_{\parallel} A_{\parallel}+{\vec \partial}_{\perp} 
{\vec A}_{\perp}-{\vec u}{\vec E}])\Psi(x)
\ee
where $\gamma=1/{\sqrt {1-{\vec u}^2}}$
and both the parallel and perpendicular components of $\vec A$ are determined with respect
to the velocity vector. As shown in \cite{Lavelle},  
Eq.(54) gives rise to the operator whose propagator is both IR-finite and UV-renormalizable
at $p_\mu =m\gamma(1, {\vec u})$.

Such a strong dependence on the exact details of the construction of the phase factor
appearing in the gauge transformation (53) indicates 
that the true electron Green function may well be quite different from 
Eq.(34). In particular, it remains to be seen whether one can at all find an alternate form
$G_{inv}^{\Gamma}(x)$ which would decay faster than the bare propagator. 
Given the intellectual appeal of the $QED_{2+1}$ picture,
such an endevour is definitely worth the effort, and 
a further investigation into this possibility is currently under way.

Should, however, the sought-after Luttinger-like behavior fail 
to occur even in the modified prototype of the electron propagator, 
one can still consider an alternative approach to the quantum disordered $d$-wave 
systems, e.g., the one that was put forward in the context of the scenario of a 
second pairing transition in the 2D superconducting phase \cite{Sachdev}.
In \cite{Jens}, apart from fully idenifying the true nature 
of this transition and its critical properties (the specific predictions
of Ref.\cite{Jens} for the critical exponents 
are roughly consistent with the recent tunneling data in $Ca$-doped 
$YBaCuO$ \cite{Deutscher}),
it was further speculated that it might be possible to extend
the effective Higgs-Yukawa theory 
of the nodal fermion excitations coupled to the fluctuations of the secondary 
order parameter of either $id_{xy}$ or $is$ symmetry well into the pseudogap phase.  
Rather than a global superconducting coherence, this 
would only require the presence of a local parent $d_{x^2-y^2}$-wave order.
If this speculation prove valid, it can provide a viable alternative to the $QED_{2+1}$-based
fits
to the ARPES data \cite{Wen,Ye},
since in the Higgs-Yukawa theory the anomalous dimension of the Dirac 
fermions is indeed $positive$ \cite{Sachdev,Jens,Reenters}.
 
To summarize, in the present paper we applied the Schwinger's
functional integral representation of the fermion amplitudes to the analysis of both the 
infrared and ultraviolet asymptotics of the conventional 
(non-gauge invariant) fermion Green function and a particular gauge-invariant amplitude (34). 

In the IR regime, this method provides a substantial improvement
with regard to the spinless Bloch-Nordsieck model or the customary  
semiclassical (eikonal) approximation, since it preserves the exact spinor
structure of the fermion amplitudes. Moreover, the intrinsic 
"exponential" form of the Schwinger's integral representation
facilitates truly non-perturbative calculations.

In the opposite, UV, regime,
the method allows one to naturally separate between the gauge invariant
and non-invariant contributions to the mass operator
and systematically compute the higher order contributions into both kinds of terms.
For a specific class of problems, including the amplitudes given by Eq.(2), 
it has a significant advantage as compared to the conventional diagrammatic technique
which is not particularly well suited for such calculations,
for the very rules of the diagrammatic expansion
turn out to be amplitude-specific and depend on a particular choice of the contour $\Gamma$
\cite{Lavelle}.

To our surprise, the previously suggested "minimal"  
form of the physical electron Green function
(34) was found to manifest a negative anomalous dimension, contrary to the
much-anticipated Luttinger liquid behavior.
The implications of this observation
were discussed, some of them pertaining to the applicability of Eq.(34) and other
to the possible alternatives to the $QED_{2+1}$-like description of the 
ARPES data in the high-$T_c$ cuprates. 

The author acknowledges valuable discussions and email communications with 
T. Appelquist, D. Dyakonov, A. Luther, N. Mavromatos, K. Martin,
J.Ng, N. Stefanis, Z. Tesanovic, and A. Yashenkin.
This research was supported by the NSF under Grant No. DMR-0071362
and also by the Aspen Center for Physics, ICTP (Trieste), and Nordita (Copenhagen) 
where part of this work was carried out.

\end{document}